\documentclass[twocolumn]{revtex4}
\usepackage{graphicx}
\usepackage{bm}
\usepackage{color}
\usepackage{amsmath}
\usepackage{natbib}

\begin{document}

\title{Spin current contribution in the spectrum of collective excitations of degenerate partially polarized spin-1/2 fermions at separate dynamics of spin-up and spin-down fermions}

\author{Pavel A. Andreev}
\email{andreevpa@physics.msu.ru}
\affiliation{Faculty of physics, Lomonosov Moscow State University, Moscow, Russian Federation.}

 \date{\today}

\begin{abstract}
The spectrum of collective excitations of degenerate partially polarized spin-1/2 fermions is considered. The spin-up fermions and the spin-down fermions are considered as different fluids. Corresponding two-fluid hydrodynamics consistent with a non-linear Pauli equation is suggested. An equation of state for the spin current caused by the distribution of particles on different energy levels is suggested for the degenerate regime, where the spin current is caused by the Pauli blocking. Spectrum of three waves is found as a solution of the hydrodynamic equations: two sound waves and one spin wave. Their spectrums are calculated for two regimes: propagation parallel and perpendicular to the direction of the equilibrium spin polarization.
\end{abstract}

\pacs{ , }


\maketitle


\section{Introduction}

Interaction and evolution of the magnetic moments are studied for different physical systems: ferromagnets \cite{Valiska PRB 15}, \cite{Simoni PRB 17}, ferrofluids\cite{Urtizberea JAP 12}, strongly interacting ultracold fermions \cite{Thywissen Sc 14}, spin-1/2 quantum plasmas \cite{Mahajan PRL 11}, \cite{Yoshida JP A 15}, finite chains of spins \cite{Tsysar CPB 15}, quantum ferrofluids of ultracold bosons \cite{Pfau Nature 16}, \cite{Verberck Nat Phys 16}, \cite{Ferrier-Barbut PRL 16} and this paper is focused on the weakly interacting spin-1/2 fermions.

In contrast with bosons, fermions have a large Fermi pressure at the zero temperature.
This relict of the pressure caused by the distribution of particles on different energy levels exists due to the Pauli blocking.
Microscopic derivation of the spin evolution equation shows that the spin current caused by the distribution of particles on different energy levels also exists \cite{MaksimovTMP 2001}, \cite{Andreev IJMP B 15}, \cite{Andreev 1510 Spin Current}.
At small temperatures it reduces to a spin current caused by the Pauli blocking. It is similar to the Fermi pressure in
the Euler equation thus it is called the Fermi spin current \cite{Andreev 1510 Spin Current}.

The explicit form of the Fermi pressure is well known for a long time. It gives an equation of state for the pressure of
degenerate spin-1/2 fermions with zero spin polarization. An account of the spin polarization in the equation of state
of the degenerate fermion gas can be easily done \cite{Margulis JETP 87}. However, the Fermi
spin current has not been considered (spin currents caused by the interaction are not
discussed here see for instance \cite{Leggett PRL 68}). Recently, an equation of state has been suggested at the analysis of
the spin imbalanced electron gas. It is based on a minimal coupling scheme which selfconsistently bind a non-linear
Pauli equation, Euler equations for spin-up and spin-down electrons and the spin evolution equation \cite{Andreev 1510 Spin Current}. The account of the
Fermi spin current dramatically modifies the spin wave spectrum \cite{Andreev 1510 Spin Current}. Moreover, the importance of the account of the
separate spin evolution at the analysis of the concentration and velocity field evolution follows from the
appearance of the spin-electron acoustic wave \cite{Andreev PRE 15 SEAW}.

This result can be applied to neutral spin-1/2 fermions considered in different aspects of the spinor quantum gases \cite{Duarte PRL 15}-\cite{Cocchi PRL 16}.
Spectrum of collective excitations in the spin imbalanced spin-1/2 fermions is calculated in this paper
considering the short-range interaction between fermions with different spin projection in the first order by
the interaction radius.

The external magnetic field creates a preferable direction in space. Therefore, we can expect anisotropy of the spectrum. Particularly, we have different spectrums for wave propagation parallel and perpendicular to external field.

Dynamics of a system with non-zero spin particles of a single species can be presented in form of the single fluid hydrodynamics \cite{Takabayasi PTP 55 b}. These hydrodynamic equations include the spin evolution equation along with equations for evolution of concentration and velocity field of all particles. However, a multi-fluid description of the system can be applied either (see for instance \cite{Szirmai PRA 12} for a three-fluid hydrodynamics of spin-1 Bose-Einstein condensates).

Suggesting model is similar to the spinor BEC \cite{Ho PRL 98}, \cite{Machida JPSJ 98}, \cite{Ueda PRA 02} (see reviews \cite{Kawaguchi Ph Rep 12} and \cite{Stamper-Kurn RMP 13} as well).
There are 2-types of the ground state in the spin-1 BEC. They are called ferromagnetic state (phase)
and polar state.In ferromagnetic phase all spins have same direction. In polar or antiferromagnetic phase  there are same numbers of spins with opposite directions or all spin-1 particles have zero projection of their spins for illustration see fig. 2 in Ref. \cite{Uchino PRA 10} and fig. 1 in Ref. \cite{Ueda AR CMP 11}. There are three branches of two types of the collective excitations in each ground state:
a Bogoliubov (sound) mode and spin modes \cite{Ho PRL 98}, \cite{Machida JPSJ 98}.
Spin-1/2 fermions also show three branches of two types of the collective excitations. The spin-3/2 and spin-5/2 Fermi gases are also under consideration in literature \cite{Eckert NJP 07}.

If we have full equilibrium spin polarization of spin-1/2 fermions they show two branches of waves: one sound wave and spin wave. If there is no spin perturbation the sound wave only exists in the system. The partial spin polarization leads to the splitting of the sound wave on two sound waves (similar splitting exists for the zero sound \cite{Andreev Landau damping}).

This paper is organized as follows. In Sec. II, the basic equations are presented in two forms: hydrodynamic form and in the form of non-linear Pauli equation. In Sec. III, the spectrum of waves propagating in ultracold neutral fermions is studied. Sec. III consists of several sunsections. In Sec. III A, the equilibrium state is considered. In Sec. III B, linearized hydrodynamic equations are presented. In Sec. III C, the spectrum of waves propagating parallel to the equilibrium spin polarization is obtained. In Sec. III D, the spectrum of waves propagating perpendicular to the equilibrium spin polarization is found. In Sec. III E, regime of zero external magnetic field is described.
In Sec. IV, a possibility of the spin acoustic solitons is discussed for the small amplitude non-linear waves propagating parallel to the equilibrium spin polarization.
In Sec. V, a summary of the obtained results is presented.

\section{Model}

We consider partial spin polarized spin-1/2 fermions.
Hence, we have the contribution of the short-range interaction between spin-up and spin-down fermions.

We suggest a minimal coupling model allowing the hydrodynamic description with a non-linear Schrodinger (Pauli) equation (NLSE). The NLSE arises as
\begin{equation}\label{SUSDN Pauli eq Non Lin} \imath\hbar\partial_{t}\psi=\biggl(-\frac{\hbar^{2}\triangle}{2m}+\hat{\pi}+\hat{g}-\mu\widehat{\mbox{\boldmath $\sigma$}} \textbf{B}\biggr)\psi \end{equation}
where $\psi$ is the spinor wave function which is a column with two elements $\{\psi_{u}, \psi_{d}\}$, subindexes $u$ and $d$ refer to the spin-up and spin-down states, $\widehat{\mbox{\boldmath $\sigma$}}$ are the Pauli matrixes,
$\hat{\pi}=\left(\begin{array}{cc} \pi_{u} & 0 \\
0 & \pi_{d} \\\end{array}\right)$ is the spinor pressure term,
which arises as a diagonal second rank spinor, where $\pi_{s}=(6\pi^{2}n_{s})^{\frac{2}{3}}\hbar^{2}/2m$ are the contribution of the Fermi pressure. It can be represented in term of the Pauli matrixes $\hat{\pi}=\pi_{u}(\hat{\textrm{I}}+\widehat{\sigma}_{z})/2+\pi_{d}(\hat{\textrm{I}}-\widehat{\sigma}_{z})/2$, where $\hat{\textrm{I}}$ is the unit second rank spinor $\hat{\textrm{I}}=\left(
\begin{array}{cc} 1 & 0 \\
0 & 1 \\\end{array}
\right)$. The short-range interaction is presented by $\hat{g}=\left(\begin{array}{cc} gn_{d} & 0 \\
0 & gn_{u} \\\end{array}\right)=g[(n_{u}+n_{d})\hat{\textrm{I}}+(n_{d}-n_{u})\widehat{\mbox{$\sigma$}}_{z}]/2$.
It gives an interspecies interaction \cite{Ashhab PRA 03}, \cite{Ashhab JLTP 05}, \cite{Junjun Xu EPL 11}.
Considering the short-range interaction we drop the contribution of terms arising in the third order on the interaction radius leading to non-local terms \cite{Andreev PRA08}, \cite{Andreev MPL B 12}, \cite{Zezyulin EPJ D 13}.
Notions spin-up and spin-down are defined relatively to the direction of equilibrium spin polarization which can be created by the external magnetic field. We choose the coordinate frame with the $z$ axis directed along the equilibrium spin polarization.
The magnetic field $\textbf{B}$ in equation (\ref{SUSDN Pauli eq Non Lin}) is the superposition of the external magnetic field $\textbf{B}_{ext}=B_{ext}\textbf{e}_{z}$, the magnetic field created by the magnetic moments of the system (the spin-spin interaction), and the effective magnetic field $\textbf{B}_{eff}$ representing the radio frequency field creating the equilibrium spin polarization.

\begin{figure}
\includegraphics[width=8cm,angle=0]{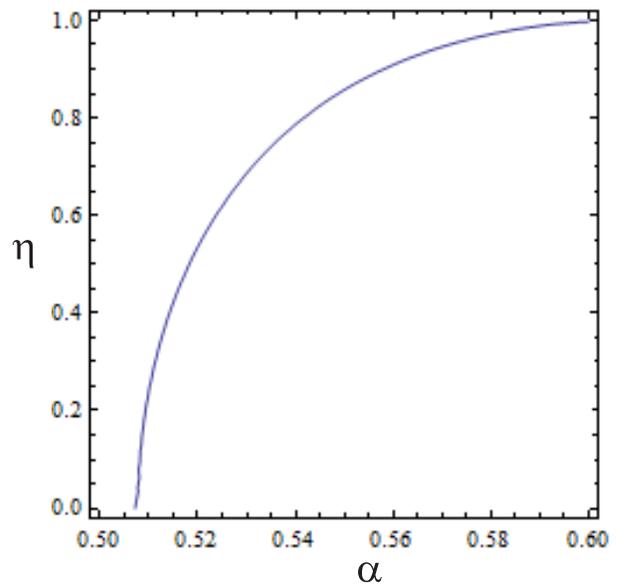}
\caption{\label{EC Feq1} (Color online) The figure shows the dependence of the spin polarization $\eta$ on the normalized scattering length $\alpha=an_{0}^{1/3}$ at the zero external magnetic field and $n_{0}=10^{14}$ cm$^{-3}$. Other parameters do not affect this dependence as it can be seen from equation (\ref{SUSDN equil state eq 1}). Further increase of the scattering length $\alpha$ remains the spin polarization equals to 1.}
\end{figure}
\begin{figure}
\includegraphics[width=8cm,angle=0]{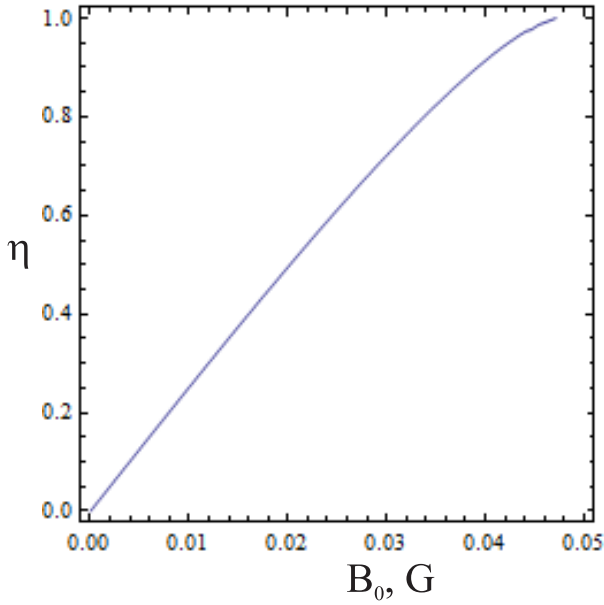}
\caption{\label{EC Feq5} (Color online) The figure shows the dependence of the spin polarization $\eta$ on the external magnetic field $B_{0}=B_{ext}+B_{eff}$ at the zero normalized scattering length $\alpha=an_{0}^{1/3}=0$, $\mu = 2\mu_{B}$, and $m = 7$ u.}
\end{figure}
\begin{figure}
\includegraphics[width=8cm,angle=0]{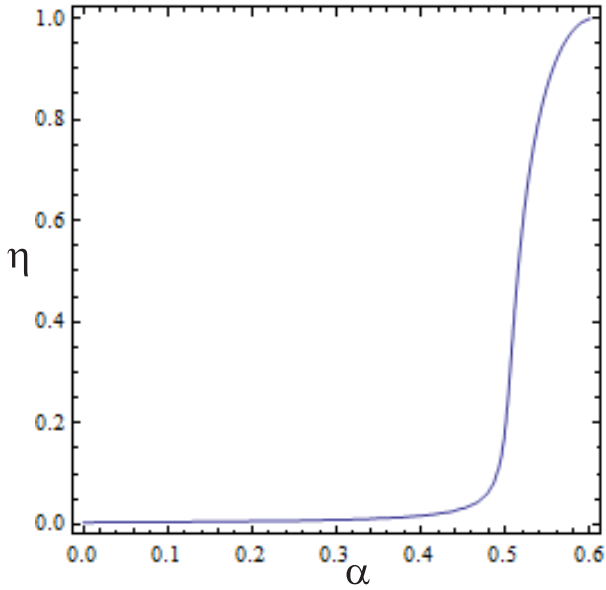}
\caption{\label{EC Feq2} (Color online) The figure shows the dependence of the spin polarization $\eta$ on the normalized scattering length $\alpha=an_{0}^{1/3}$ at $B_{0} = 0.001$ G, $n_{0}=10^{14}$ cm$^{-3}$, $\mu = 2\mu_{B}$, and $m = 1$ u, where $u$ is the unified atomic mass unit (Dalton). The figure is obtained as a solution of equation (\ref{SUSDN equil state eq 3}).}
\end{figure}
\begin{figure}
\includegraphics[width=8cm,angle=0]{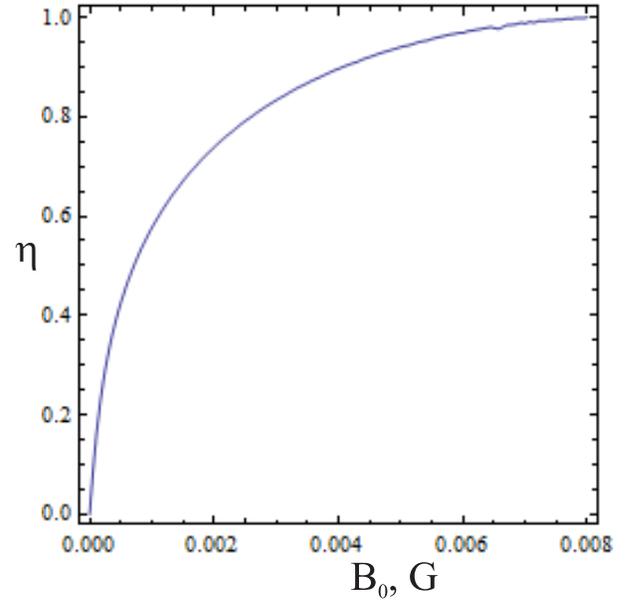}
\caption{\label{EC Feq4} (Color online) The figure shows the dependence of the spin polarization $\eta$ on the magnetic field $B_{0}$ at fixed normalized scattering length $\alpha=an_{0}^{1/3}=0.5$ (it corresponds to $a=1.1\times10^{-5}$ cm), $\mu = 2\mu_{B}$, and $m = 7$ u. Further increase of the magnetic field remains the spin polarization equals to $\eta=1$.}
\end{figure}
\begin{figure}
\includegraphics[width=8cm,angle=0]{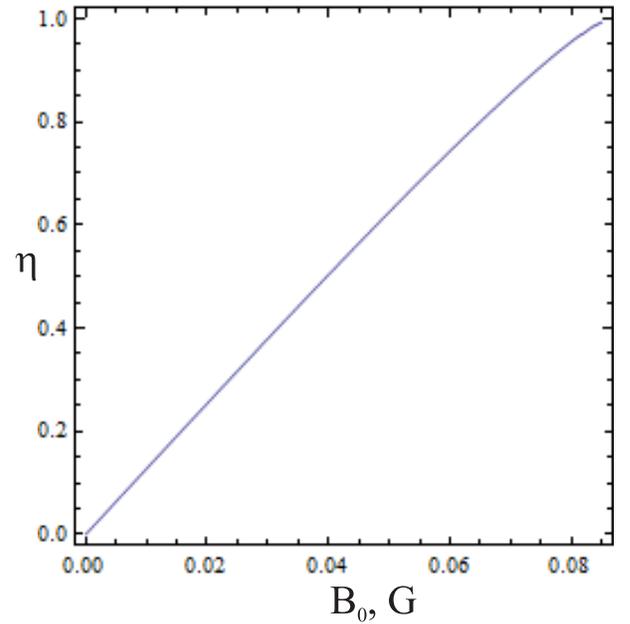}
\caption{\label{EC Feq6} (Color online) The figure shows the dependence of the spin polarization $\eta$ on the magnetic field $B_{0}$ at negative normalized scattering length $\alpha=an_{0}^{1/3}=-0.5$ (it corresponds to $a=-1.1\times10^{-5}$ cm), $\mu = 2\mu_{B}$, and $m = 7$ u.}
\end{figure}

The continuity equations corresponding to the minimal coupling model (\ref{SUSDN Pauli eq Non Lin}) appears as follows
\begin{equation}\label{SUSDN cont eq spin UP}
\partial_{t}n_{s}+\nabla(n_{s}\textbf{v}_{s})=\pm\frac{\mu}{\hbar}(S_{x}B_{y}-S_{y}B_{x}), \end{equation}
where we have applied $S_{x}$ and $S_{y}$ for mixed combinations of $\psi_{u}$ and $\psi_{d}$, $\pm$ means $+$ at $s=u$ and $-$ at $s=d$, $n_{s}=\psi_{s}^{*}\psi_{s}$, $S_{\alpha}=\psi^{\dagger}\widehat{\sigma}_{\alpha}\psi$, with $^{*}$ is the complex conjugation and $^{\dagger}$ the hermitian conjugation.

The velocity fields presented in the continuity equations (\ref{SUSDN cont eq spin UP}) are satisfy the following
Euler equations
$$mn_{s}(\partial_{t}+\textbf{v}_{s}\nabla)\textbf{v}_{s}+\nabla p_{s}$$
\begin{equation}\label{SUSDN Euler eq spin UP and D} -\frac{\hbar^{2}}{4m}n_{s}\nabla\Biggl(\frac{\triangle n_{s}}{n_{s}}-\frac{(\nabla n_{s})^{2}}{2n_{s}^{2}}\Biggr)=\textbf{F}_{g,s}+\textbf{F}_{S,s},\end{equation}
where $\textbf{F}_{g,s}$ is the force field of short-range interaction: $\textbf{F}_{g,u}=-gn_{u}\nabla n_{d}$, $\textbf{F}_{g,d}=-gn_{d}\nabla n_{u}$, $\textbf{F}_{S,s}$ is the force field of spin-spin interaction, its explicit form arises as
$$F_{S,s}^{\alpha}=\pm\mu n_{s}\partial^{\alpha} B_{z} +\frac{\mu}{2}(S_{x}\partial^{\alpha} B_{x}+S_{y}\partial^{\alpha} B_{y})$$
\begin{equation}\label{SUSDN force ss int} \pm\frac{m\mu}{\hbar}((J_{(M)}^{x\alpha}-S_{x}v_{s}^{\alpha})B_{y}-(J_{(M)}^{y\alpha}-S_{y}v_{s}^{\alpha})B_{x}),\end{equation}
with the following explicit form of the substantial (material, convective) and quantum parts of spin currents
\begin{equation}\label{SUSDN Spin current x} J_{(M)}^{\alpha\beta}=\frac{1}{2}(v_{u}^{\beta}+v_{d}^{\beta})S^{\alpha}-\varepsilon^{\alpha\gamma z}\frac{\hbar}{4m} \biggl(\frac{\partial^{\beta} n_{u}}{n_{u}}-\frac{\partial^{\beta} n_{d}}{n_{d}}\biggr)S^{\gamma}, \end{equation}
and pressures
\begin{equation}\label{SUSDN EqState P partial} p_{s}=\frac{2}{5}\pi_{s}n_{s}.\end{equation}

The generalized Bloch equation describes the spin evolution which is related to evolution of both species simultaneously:
$$n(\partial_{t}+\textbf{v}\nabla) \frac{\textbf{S}}{n}+\frac{(6\pi^{2})^{\frac{2}{3}}\hbar}{m}(n_{u}^{\frac{2}{3}}-n_{d}^{\frac{2}{3}})[\textbf{S},\textbf{e}_{z}]$$
\begin{equation}\label{SUSDN eq of magnetic moments evol} -\frac{\hbar}{2m}\partial^{\beta}[\textbf{S},\partial^{\beta}\frac{\textbf{S}}{n}] =\frac{2\mu}{\hbar}[\textbf{S},\textbf{B}]+\frac{g}{\hbar}(n_{u}-n_{d})[\textbf{S},\textbf{e}_{z}],\end{equation}
where $[\textbf{a},\textbf{b}]$ is the vector product of vectors $\textbf{a}$ and $\textbf{b}$, $n=n_{u}+n_{d}$ is the full concentration, and $\textbf{v}=(n_{u}\textbf{v}_{u}+n_{d}\textbf{v}_{d})/n$ is the velocity field of whole system. The Fermi spin current is presented by the second term.

The hydrodynamic equations (\ref{SUSDN cont eq spin UP})-(\ref{SUSDN eq of magnetic moments evol}) are coupled with the Maxwell magneto--static equations which  are
$\nabla\textbf{B}=0$, $\nabla\times \textbf{B}=4\pi\mu\nabla\times \textbf{S}$,
where $\textbf{S}=\{S_{x}, S_{y}, n_{u}-n_{d}\}$ is the three vector of spin density.

It is well-known that hydrodynamic equations do not include the zero-sound which can be included by a kinetic model \cite{Landau v10}.

The hydrodynamic model presented by equations (\ref{SUSDN cont eq spin UP})-(\ref{SUSDN eq of magnetic moments evol}) is similar to the separate spin evolution quantum hydrodynamics developed for electron gas \cite{Andreev PRE 15 SEAW}.

It is well-known that the hydrodynamic form of equations (which sometimes can be represented as the non-linear Schrodinger equation) for fermions is justified
for two phases: the superfluid phase and the normal phase for the strongly interacting regime.
For instance, a phenomenological hydrodynamic model of one dimensional motion of cold fermions in the Boussinesque approximation with dissipation is considered in Ref. \cite{Kulkarni PRA 12} for the study of weakly nonlinear wave perturbations in such systems.
However, in this paper, we consider fermions in the normal phase
for weakly interacting regime. Hence, the  basic equations require a justification. Disregarding the spin-spin interaction and the separate spin evolution, a
hydrodynamic model of fermions with short range interaction is derived for weakly interacting regime \cite{Andreev PRA08}. More precisely, it is shown that a hydrodynamic form
of equations of motion can be always found for fermions of a chosen species. However, the explicit forms of the pressure term and the force field require the
specification of the regime and appear to be a problem in most of the cases. The pressure and the force field are approximately found in \cite{Andreev PRA08} for weakly
interacting fermions. The well-known Fermi pressure is considered as the equation of state in this regime. The short range interaction between the fermions
having same spin spin projection is equal to zero in the first order on the interaction radius. It happens due to the antisymmetry of the wave function
relatively the permutation of particles. However, a nonzero force field appears in the next order. The third order on the interaction radius approximation leads
to the nonlocal terms describing the short range interaction presented by equations (53), (59) in Ref. \cite{Andreev PRA08} and equation (7) in Ref. \cite{Zezyulin EPJ D 13}. Interaction between spin-up and spin-down fermions can be
considered as an interspecies interaction. It is also considered in Ref. \cite{Andreev PRA08}, where bosons and fermions are considered as an example of two species, but the
result is relevant for other pairs of species. Hence, this result is used in equations (\ref{SUSDN Pauli eq Non Lin}) and (\ref{SUSDN Euler eq spin UP and D}).

\subsection{On the possibility of the hydrodynamic description of fermions}

System of fermions can be described by the many-particle wave function (wave spinor) $\Psi(R,t)=\Psi(\textbf{r}_{1}, ..., \textbf{r}_{N}, t)$.
It allows to construct collective variables as the quantum mechanical average of required operators.
One of the simplest collective variables known from hydrodynamics is the concentration of particles which can be defined as averaging of operator $\sum_{i=1}^{N}\delta(\textbf{r}-\textbf{r}_{i})$.
It gives the following definition
\begin{equation}\label{SUSDN concentration mp}
n_{f}(\textbf{r},t)=\int \Psi^{+}(R,t)\sum_{i=1}^{N}\delta(\textbf{r}-\textbf{r}_{i})\Psi(R,t)dR.\end{equation}
Evolution of concentration is determined by the Hamiltonian via the time evolution of the wave function.

Next, we can go further and split wave spinor $\Psi(R,t)$ on two column $\Psi_{u}(R,t)$ and $\Psi_{d}(R,t)$ which are related to the spin-up and spin-down states of each particle. 
It allows us to introduce the partial concentrations for spin-1/2 fermions
\begin{equation}\label{SUSDN concentration mp}
n_{fs}(\textbf{r},t)=\int \Psi_{s}^{+}(R,t)\sum_{i=1}^{N}\delta(\textbf{r}-\textbf{r}_{i})\Psi_{s}(R,t)dR.\end{equation}
Evolution of these functions reveals in the hydrodynamic equations which can be represented in non-linear Pauli equation introduced above (\ref{SUSDN Pauli eq Non Lin}) in accordance with Refs. \cite{Andreev 1510 Spin Current}, \cite{Andreev PRE 15 SEAW}, \cite{Andreev Landau damping}.

\section{Spectrum of collective excitations: Linear regime}

\subsection{Equilibrium state}

Consider an equilibrium state. To this end, the wave function is presented in the following form $\psi_{s}=A_{s}e^{-\imath\mu_{ch} t/\hbar}$, where $\mu_{ch}$ is the chemical potential and $A_{s}$ are constant which do not depend on $\textbf{r}$ and $t$. Hence, we have $\triangle\psi_{s}=0$ in the equilibrium.
If there is no short range interaction and external magnetic field, equation (\ref{SUSDN Pauli eq Non Lin}) reduces to $\imath\hbar\partial_{t}\psi_{s}=\hat{\pi}\psi_{s}$. Substituting $\psi_{s}$ explicitly, we find $\mu_{ch}\psi_{s}=\hat{\pi}\psi_{s}$. Consequently, we obtain $\mu_{ch}=\pi_{u}=\pi_{d}$ and $n_{u}=n_{d}=n_{0}/2$.
If there is no external magnetic field at zero temperature all fermions occupy quantum states with lower energies. All states below the Fermi energy $\varepsilon_{Fe}=(3\pi^{2}n)^{\frac{2}{3}}\hbar^{2}/2m=\pi_{s}(n_{s}=n_{0}/2)$ are occupied due to the Pauli blocking. So, there is no spin polarization and we have the polar equilibrium state since we have equal numbers of fermions in states with opposite spin projections. It is similar to the polar state described in \cite{Ho PRL 98}, \cite{Ueda AR CMP 11}.

The external magnetic field creates the spin imbalance since it transfers energy to some spin-down fermions to change their direction to spin-up states (it is assumed that the Lande factor is positive).
Obviously, the external magnetic field creates the spin polarization, but we consider the short range interaction with no external magnetic field. It gives the following equations $\mu_{ch}=\pi_{u}+gn_{d}$ and $\mu_{ch}=\pi_{d}+gn_{u}$. Consequently, we find $\pi_{u}+gn_{d}=\pi_{d}+gn_{u}$. It leads to
\begin{equation}\label{SUSDN equil state eq 1}(n_{u}^{\frac{1}{3}}-n_{d}^{\frac{1}{3}})\biggl(n_{u}^{\frac{1}{3}}+n_{d}^{\frac{1}{3}}-\frac{8\pi a}{(6\pi^{2})^{\frac{2}{3}}}(n_{u}^{\frac{2}{3}}+n_{u}^{\frac{1}{3}}n_{d}^{\frac{1}{3}}+n_{d}^{\frac{2}{3}})\biggr)=0.\end{equation}
There are two solutions. One of them describes the polar phase with $n_{u}=n_{d}$.
Fig. \ref{EC Feq1} confirms that equation (\ref{SUSDN equil state eq 1}) has a nontrivial solution.
This equation has the second solution for the repulsive short range interaction $a>0$. There is a narrow interval for the second solution.
A solution with the experimentally achievable concentrations $n_{0}=n_{0u}+n_{0d}\sim 10^{12}\div10^{14}$ cm$^{-3}$ exists at specific scattering length $a\sim10^{-6}$ cm, which are realistic scattering length. Therefore, the partially spin polarized state can be realized for spin-1/2 fermions due to the repulsive short range interaction between fermions with different spin projections.

Presence of the magnetic field or the radio frequency field creating equilibrium spin polarization, so it can be describe described as an effective equilibrium magnetic field $\textbf{B}_{eff}$, changes both possible equilibrium states.
It creates the polarization in case of the zero initial polarization or increases the spin polarization created by the repulsive short range interaction. The attractive short range interaction decreases the spin polarization created by the external or effective magnetic field.
Consider it in more details. To this end, rewrite the Pauli equation (\ref{SUSDN Pauli eq Non Lin}) for this equilibrium state: $\mu_{ch}\psi_{u}=(\pi_{u}+gn_{d}-\mu B_{z})\psi_{u}$ and
$\mu_{ch}\psi_{d}=(\pi_{d}+gn_{u}+\mu B_{z})\psi_{d}$, where $B_{z}\rightarrow B_{0}=B_{ext}+B_{eff}$. Equating the different forms of the chemical potential $\mu_{ch}$ obtained from these two equations we
find the following relation between $n_{u}$ and $n_{d}$:
\begin{equation}\label{SUSDN equil state eq 2} \pi_{u}-\pi_{d}=g(n_{u}-n_{d})+2\mu B_{0} \end{equation}
containing two parameters $g$ and $B_{0}$.

Equation (\ref{SUSDN equil state eq 2}) can be rewritten in terms of the spin polarization:
\begin{equation}\label{SUSDN equil state eq 3} (1+\eta)^{\frac{2}{3}}-(1-\eta)^{\frac{2}{3}}-\eta\frac{8\pi an_{0}^{\frac{1}{3}}}{(3\pi^{2})^{\frac{2}{3}}}  -\frac{4m\mu B_{0}}{(3\pi^{2})^{\frac{2}{3}}n_{0}^{\frac{2}{3}}\hbar^{2}}=0. \end{equation}

Consider a characteristic concentration $n_{0}=10^{14}$ cm$^{-3}$ and solve numerically equation (\ref{SUSDN equil state eq 1}) (regime of zero magnetic field). The solution is shown in Fig. \ref{EC Feq1}. It demonstrates that change of the scattering length near $a\approx100$ nm allows to reach any value of spin polarization $\eta\in [0,1]$.

Include the magnetic field and solve equation (\ref{SUSDN equil state eq 3}) numerically for different regimes. A solution for zero short range interaction scattering length $a=0$, relatively small mass $m=7$ u, $\mu=2\mu_{B}$ is shown in Fig. \ref{EC Feq5}, where $u$ is the unified atomic mass unit (Dalton). It demonstrates that relatively small magnetic field $B_{0}\sim0.05$ G can create the full spin polarization $\eta=1$. The full spin polarization can be reached at smaller magnetic field for atoms of larger masses and larger magnetic moments. Influence of small magnetic field $B_{0}=0.001$ G on the dependence of the spin polarization on the scattering length is demonstrated in Fig. \ref{EC Feq2}.

As it is mentioned above, the repulsive short range interaction make it easier to create the spin polarization by the magnetic field (see Fig. \ref{EC Feq4}) and the attractive short range interaction require larger magnetic field to create the required spin polarization (see Fig. \ref{EC Feq6}).

Relatively large mass of particles (in compare with the electron) and small concentrations (in compare with condense matter physics $n_{0}\sim10^{18}\div10^{22}$ cm$^{-3}$) leads to small Fermi energy $\varepsilon_{Fe}=0.62\times10^{-20}$ egr, where $m=1$ u, $n_{0}=10^{14}$ cm$^{-3}$. Hence, the Zeeman energy $\mu B_{0}=0.9\times10^{-20}$ erg at $B_{0}=1$ G overcome the Fermi energy creating conditions for the full spin polarization.

\subsection{Small perturbations}

Next, the small perturbations of an equilibrium state are considered. The equilibrium state is described by the concentrations $n_{0u}$, $n_{0d}$, the external magnetic field $\textbf{B}_{0}=B_{0}\textbf{e}_{z}$, and the spin density projection on the magnetic field direction $S_{0z}=n_{0u}-n_{0d}$. Other quantities like the velocity fields $\textbf{v}_{u}$, $\textbf{v}_{d}$, the transverse spin density projections $S_{x}$, $S_{y}$ are equal to zero. We consider the perturbations in form of plane monochromatic wave such as a perturbation of each quantity can be presented in the following form $\delta f=F \exp(-\imath\omega t+\imath \textbf{k}\textbf{r})$.

\begin{figure}
\includegraphics[width=8cm,angle=0]{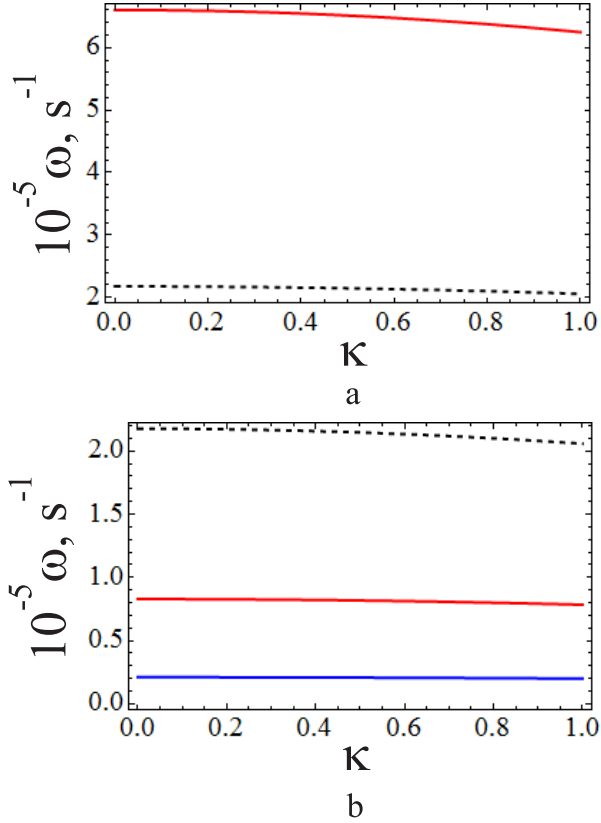}
\caption{\label{SEPAWOP F1} (Color online)
The figure (a) shows the dispersion dependence of spin waves propagating parallel to the external magnetic field. The frequency is plotted as a function of dimensionless wave vector $\kappa=k/n_{0}^{1/3}$. The figure shows the dispersion dependence for different spin polarization. The upper (lower) curve is obtained for $\eta=0.3$ ($\eta=0.1$). Other parameters are $n_{0}=10^{14}$ cm$^{-3}$, $\mu=2\mu_{B}$, $m=6$ u, $g=4\pi \hbar^{2}a/m$, $\alpha=-0.4$, where $\mu_{B}$ is the Bohr magneton.
The figure (b) shows the dispersion dependence for different mass of particles.
The upper (middle, lower) curve is obtained for $m=6$ u ($m=16$ u, $m=66$ u) at $\eta=0.1$. Other parameters are $n_{0}=10^{14}$ cm$^{-3}$, $\mu=2\mu_{B}$, $\alpha=-0.4$.}
\end{figure}
\begin{figure}
\includegraphics[width=8cm,angle=0]{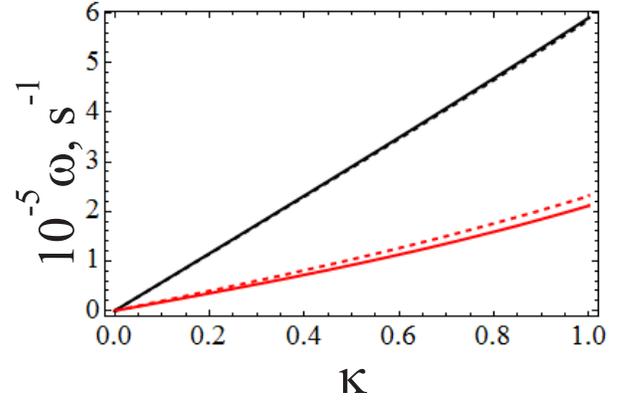}
\caption{\label{SEPAWOP Facw1} (Color online) The figure shows two sound waves which are presented by equation (\ref{SUSDN spectrum parallel longitud part}). Each of them is presented for two different spin polarizations. Continuous lines show the dispersion dependencies for $\eta=0.3$. The dashed lines present the sound waves at $\eta=0.1$ (the black dashed line almost coincide with the black continuous line). The figure shows that the splitting between the dispersion dependencies of the sound waves increases with the increase of the spin polarization.Other parameters are the following: $m=6$ u, $n_{0}=10^{14}$ cm$^{-3}$, $\mu=2\mu_{B}$, $g=4\pi \hbar^{2}a/m$, $\mid\alpha\mid=\mid a\mid n_{0}^{1/3}=0.4$, $B_{0}=10^{-3}$ G for $\eta=0.1$ and $B_{0}=3.1\times10^{-3}$ G for $\eta=0.3$.}
\end{figure}

Consider the linearized set of separate spin evolution quantum hydrodynamic equations (\ref{SUSDN cont eq spin UP})-(\ref{SUSDN eq of magnetic moments evol})
\begin{equation}\label{SUSDN cont eq spin lin}
\partial_{t}\delta n_{s}+n_{0s}\nabla\delta\textbf{v}_{s}=0, \end{equation}
\begin{equation}\label{SUSDN Euler eq spin UP and D lin}m\partial_{t}\delta\textbf{v}_{s}+\frac{\nabla \delta p_{s}}{n_{0s}}-\frac{\hbar^{2}}{4m}\frac{\nabla\triangle\delta n_{s}}{n_{0s}}=-g\nabla\delta n_{s'}\pm\mu \nabla\delta B_{z},\end{equation}
and
$$\partial_{t} \delta\textbf{S}+\frac{(6\pi^{2})^{\frac{2}{3}}\hbar}{m}(n_{0u}^{\frac{2}{3}}-n_{0d}^{\frac{2}{3}})[\delta\textbf{S},\textbf{e}_{z}]-\frac{\hbar}{2m}[\textbf{S}_{0},\triangle\frac{\delta\textbf{S}}{n_{0}}]$$
\begin{equation}\label{SUSDN eq of magnetic moments evol lin} =\frac{2\mu}{\hbar}([\delta\textbf{S},\textbf{B}_{0}]+[\textbf{S}_{0},\delta\textbf{B}])+\frac{g}{\hbar}(n_{0u}-n_{0d})[\delta\textbf{S},\textbf{e}_{z}],\end{equation}
which allows to find the spectrum of small perturbations, $s'\neq s$. They are coupled with the following linearized equations of field
\begin{equation}\label{SUSDN field eq lin}
\begin{array}{cc}   \nabla\times\delta \textbf{B}=4\pi\mu\nabla\times\delta \textbf{S}, & \nabla \delta \textbf{B}=0.
\end{array}\end{equation}

Below we consider different types of waves. Let us discuss some general properties of the sound waves and the spin waves.
Sound waves have linear spectrum $\omega\sim k$ which can be modified by the quantum Bohm potential, so $\omega^{2}\sim k^{2}+ak^{4}$. In both cases the frequency is equal to zero at the zero wave vector $\omega(k=0)=0$. Sound waves can exist as density wave if there is no spin density perturbations. Existence of two spin projections for spin-1/2 fermions leads to two concentrations for particles with spin-up $n_{u}$ and particles with spin-down $n_{d}$. Increase of the species number can increase the number of sound waves. In-phase evolution of $n_{u}$ and $n_{d}$ gives the density wave. Different evolution of $n_{u}$ and $n_{d}$ gives the evolution of the spin projection on the direction of external field $\delta S_{z}=\delta n_{u}-\delta n_{d}$ with approximately linear spectrum, so we refer to it as the spin acoustic wave. In terms of the separate spin evolution this wave is a density wave since $n_{u}$ and $n_{d}$ are independent variables while $S_{z}$ is not an independent variable in the model.

The spin waves have nonzero frequency at the zero wave vector $\omega(0)\neq0$. In the simplest case, the frequency $\omega(0)$ is equal to the cyclotron frequency $\Omega$ describing the precession of the single magnetic moment in the external magnetic field.
Different many particle effects such as the Fermi spin current and the spin-spin interaction can change this frequency. The frequency dependance on the wave vector usually is quadratic $\omega\sim1+bk^{2}$, where the term proportional to $k^{2}$ caused by the quantum Bohm potential.

\begin{figure}
\includegraphics[width=8cm,angle=0]{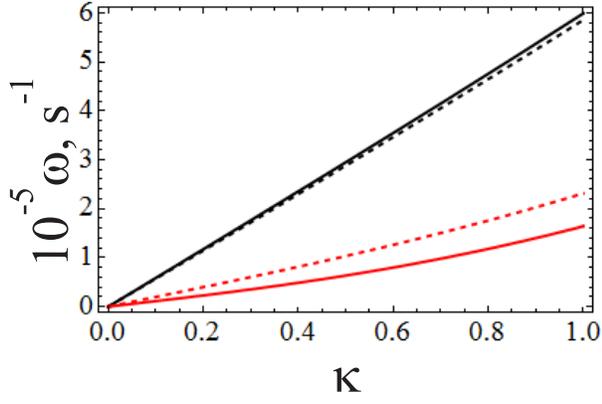}
\caption{\label{SEPAWOP Facw2} (Color online) The figure shows two sound waves which are presented by equation (\ref{SUSDN spectrum parallel longitud part}). Each of them is presented for two different spin polarizations. One of the chosen spin polarizations is larger than in Fig. \ref{SEPAWOP Facw1}. Continuous lines show the dispersion dependencies for $\eta=0.5$. The dashed lines present the sound waves at $\eta=0.1$. Other parameters are the following $m=6$ u, $n_{0}=10^{14}$ cm$^{-3}$, $\mid\alpha\mid=0.4$ and coincide with parameters in Fig. \ref{SEPAWOP Facw1}.}
\end{figure}

\begin{figure}
\includegraphics[width=8cm,angle=0]{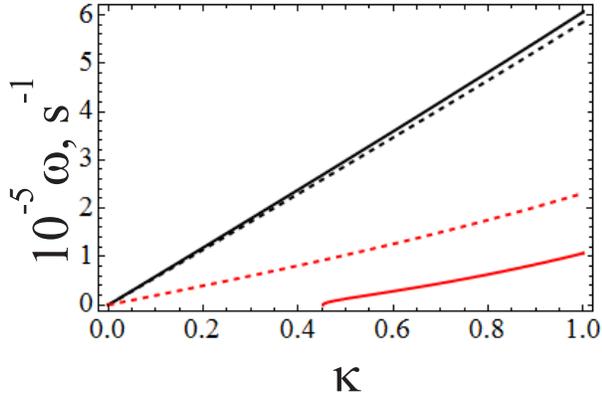}
\caption{\label{SEPAWOP Facw3} (Color online) The figure shows two sound waves which are presented by equation (\ref{SUSDN spectrum parallel longitud part}) in another regime of the spin polarization. Each solution is presented for two different spin polarizations. Continuous lines show the dispersion dependencies for $\eta=0.62$. The dashed lines present the sound waves at $\eta=0.1$. Other parameters are the following $m=6$ u, $n_{0}=10^{14}$ cm$^{-3}$, $\mid\alpha\mid=0.4$ and coincide with parameters in Fig. \ref{SEPAWOP Facw1}.}
\end{figure}

\begin{figure}
\includegraphics[width=8cm,angle=0]{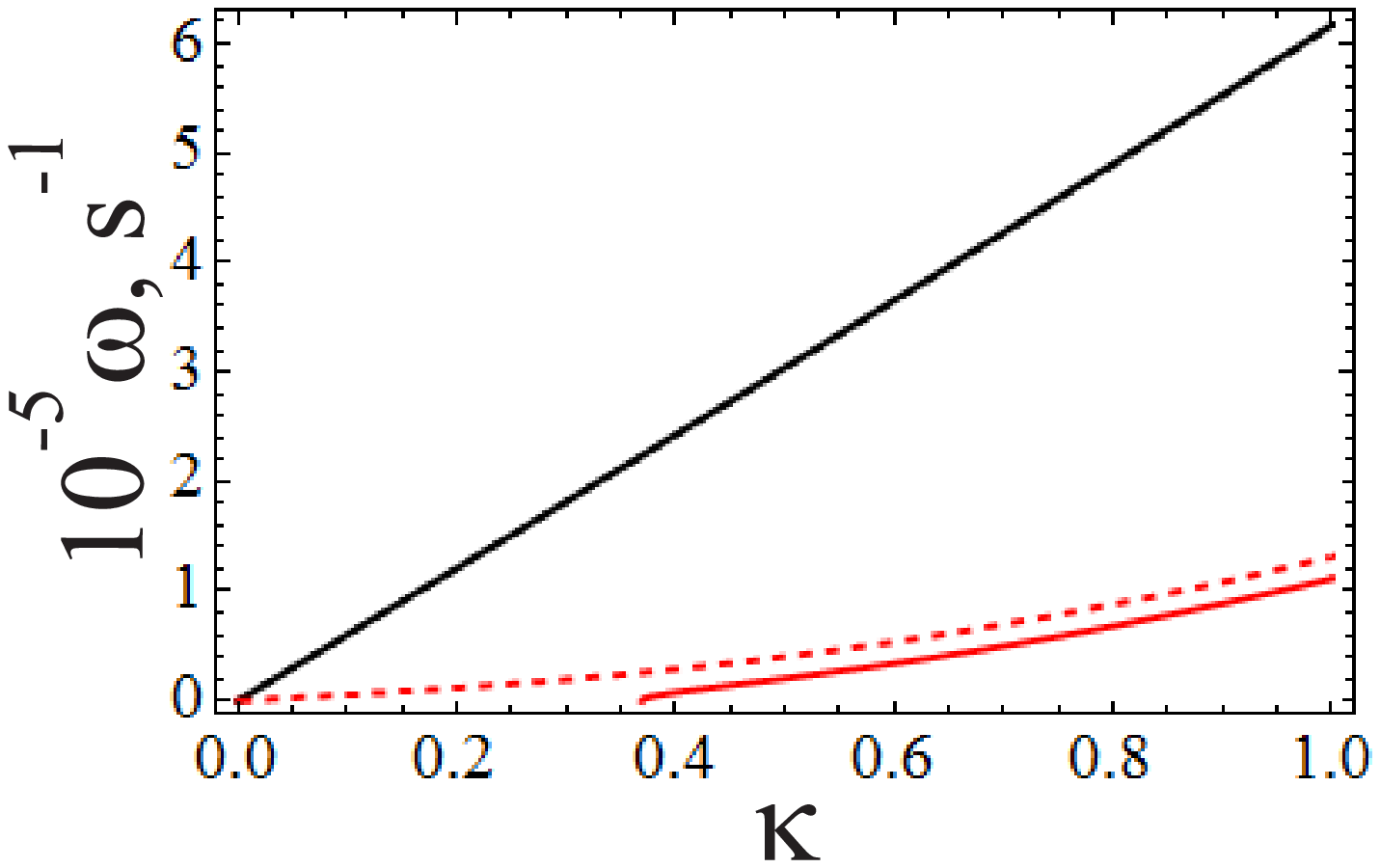}
\caption{\label{SEPAWOP Facw4} (Color online) The figure shows two sound waves which are presented by equation (\ref{SUSDN spectrum parallel longitud part}) for the following spin polarizations $\eta=0.3$ (continuous lines) and $\eta=0.1$ (dashed lines). This figure presents results for a larger scattering length $\mid\alpha\mid=\mid a\mid n_{0}^{1/3}=0.5$. Other parameters are the following $m=6$ u, $n_{0}=10^{14}$ cm$^{-3}$.}
\end{figure}

\begin{figure}
\includegraphics[width=8cm,angle=0]{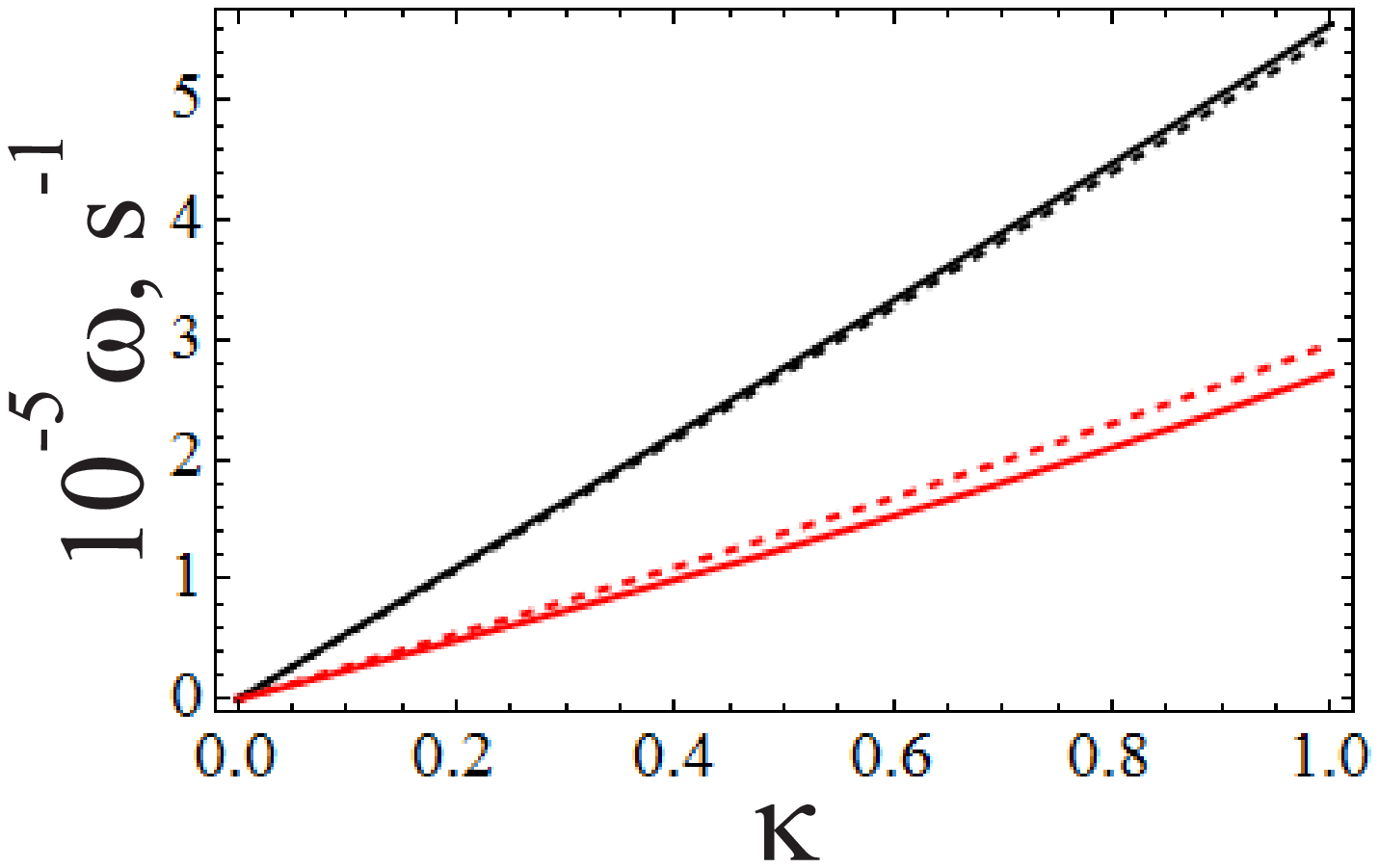}
\caption{\label{SEPAWOP Facw5} (Color online) The figure shows two sound waves which are presented by equation (\ref{SUSDN spectrum parallel longitud part}) for the following spin polarizations $\eta=0.3$ (continuous lines) and $\eta=0.1$ (dashed lines). This figure presents results for the following scattering length $\mid\alpha\mid=\mid a\mid n_{0}^{1/3}=0.3$. Other parameters are the following $m=6$ u, $n_{0}=10^{14}$ cm$^{-3}$.}
\end{figure}

\begin{figure}
\includegraphics[width=8cm,angle=0]{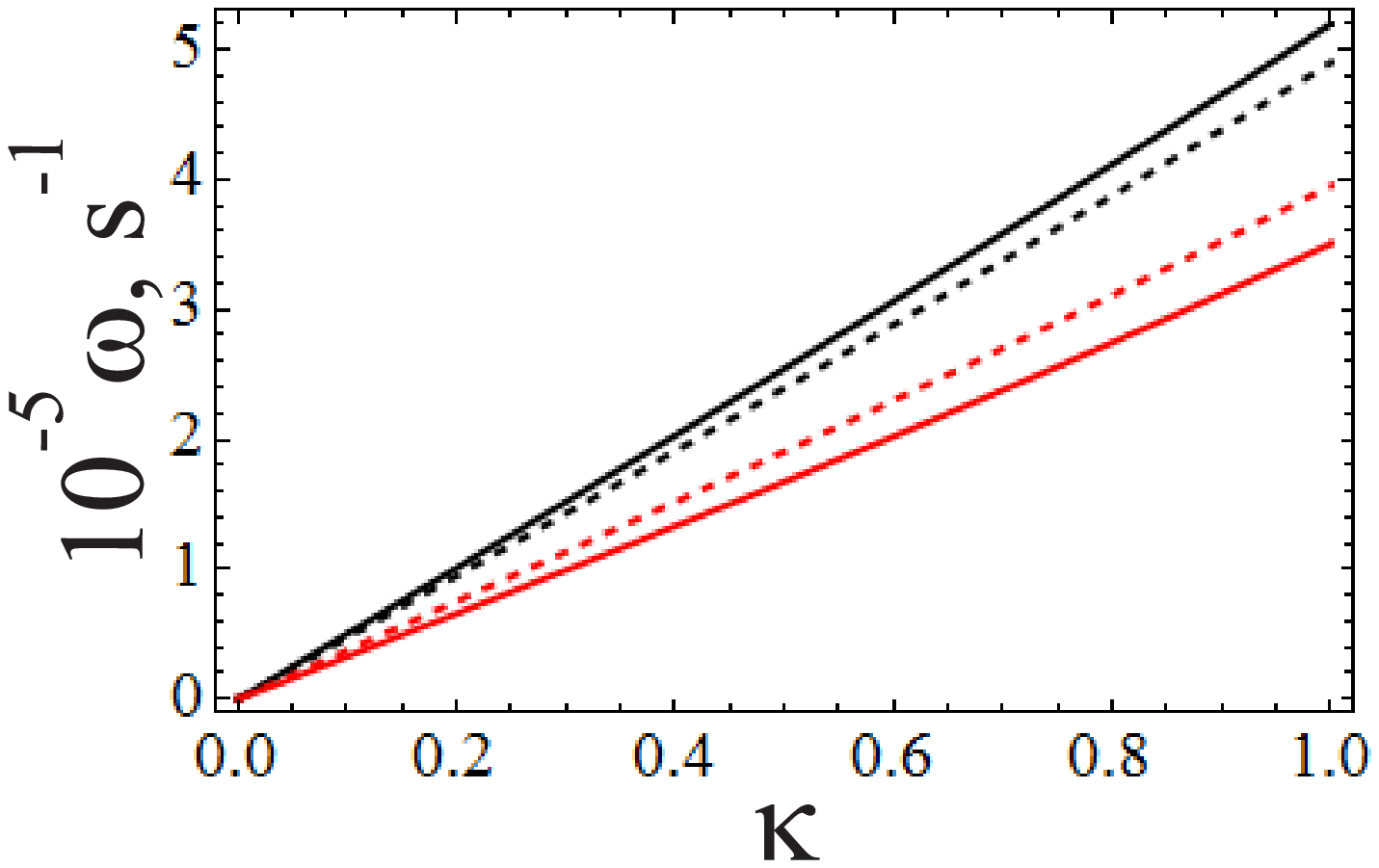}
\caption{\label{SEPAWOP Facw6} (Color online) The figure shows two sound waves which are presented by equation (\ref{SUSDN spectrum parallel longitud part}) for the following spin polarizations $\eta=0.3$ (continuous lines) and $\eta=0.1$ (dashed lines). This figure presents results for the following scattering length $\mid\alpha\mid=\mid a\mid n_{0}^{1/3}=0.1$. Other parameters are the following $m=6$ u, $n_{0}=10^{14}$ cm$^{-3}$.}
\end{figure}

\begin{figure}
\includegraphics[width=8cm,angle=0]{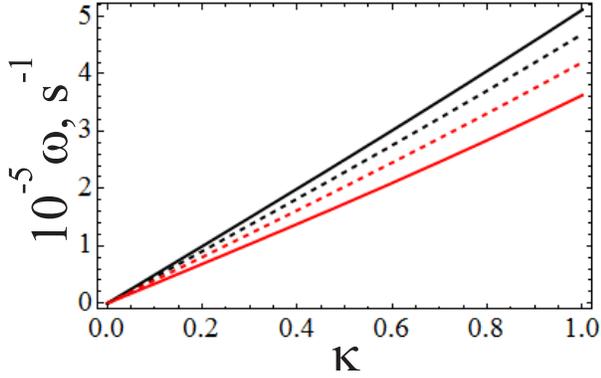}
\caption{\label{SEPAWOP Facw7} (Color online) The figure shows two sound waves which are presented by equation (\ref{SUSDN spectrum parallel longitud part}) for the following spin polarizations $\eta=0.3$ (continuous lines) and $\eta=0.1$ (dashed lines). This figure presents results for the following scattering length $\mid\alpha\mid=\mid a\mid n_{0}^{1/3}=0.01$. Other parameters are the following $m=6$ u, $n_{0}=10^{14}$ cm$^{-3}$.}
\end{figure}

\subsection{Propagation parallel to the equilibrium spin polarization}

At the propagation parallel to the anisotropy direction created by the equilibrium spin density and the external magnetic field we choose the wave vector in the following form $\textbf{k}=\{0,0,k\}$.

In this regime equations (\ref{SUSDN cont eq spin lin})-(\ref{SUSDN field eq lin}) are simplified to the following form:
\begin{equation}\label{SUSDN cont eq spin lin par}
\omega\delta n_{s}=n_{0s}k\delta v_{sz}, \end{equation}
for the continuity equation,
\begin{equation}\label{SUSDN Euler eq spin UP and D lin par}-\omega\delta v_{sz}+\frac{k \delta p_{s}}{mn_{0s}}+\frac{\hbar^{2}k^{3}}{4m^{2}}\frac{\delta n_{s}}{n_{0s}}=-\frac{gk}{m}\delta n_{s'},\end{equation}
for the Euler equation,
$$-\imath\omega \delta S_{x}+\frac{(6\pi^{2})^{\frac{2}{3}}\hbar}{m}(n_{0u}^{\frac{2}{3}}-n_{0d}^{\frac{2}{3}})\delta S_{y}-\frac{\hbar k^{2}}{2m}\frac{S_{0}}{n_{0}}\delta S_{y}$$
\begin{equation}\label{SUSDN eq of magnetic moments evol lin par x} -\frac{2\mu}{\hbar}\delta S_{y}B_{0}-\frac{g}{\hbar}(n_{0u}-n_{0d})\delta S_{y}=-\frac{8\pi\mu^{2}}{\hbar}S_{0}\delta S_{y},\end{equation}
for the $x$-projection of the spin evolution equation,
$$-\imath\omega \delta S_{y}-\frac{(6\pi^{2})^{\frac{2}{3}}\hbar}{m}(n_{0u}^{\frac{2}{3}}-n_{0d}^{\frac{2}{3}})\delta S_{x}+\frac{\hbar k^{2}}{2m}\frac{S_{0}}{n_{0}}\delta S_{x}$$
\begin{equation}\label{SUSDN eq of magnetic moments evol lin par y} +\frac{2\mu}{\hbar}\delta S_{x}B_{0}+\frac{g}{\hbar}(n_{0u}-n_{0d})\delta S_{x}=\frac{8\pi\mu^{2}}{\hbar}S_{0}\delta S_{x},\end{equation}
for the $y$-projection of the spin evolution equation,
and
\begin{equation}\label{SUSDN field eq lin par}\begin{array}{c}
                                            \textbf{e}_{x}(\delta B_{y}-4\pi\mu\delta S_{y}) -\textbf{e}_{y}(\delta B_{x}-4\pi\mu\delta S_{x})=0, \\
                                            \delta B_{z}=0,
                                          \end{array}
\end{equation}
for the Maxwell equations, with
$\delta v_{sx}=\delta v_{sy}=0$, $\delta p_{s}=(6\pi^{2}n_{0s})^{2/3}\delta n_{s}/3m$, where it is included that the last term in the Euler equation is equal to zero since $\delta B_{z}=0$ in accordance with the Maxwell equations (\ref{SUSDN field eq lin par}). The right-hand side of the spin evolution equations (\ref{SUSDN eq of magnetic moments evol lin par x}), (\ref{SUSDN eq of magnetic moments evol lin par y}) contain perturbations of the magnetic field expressed via the spin density in accordance with the Maxwell equations (\ref{SUSDN field eq lin par}).

The continuity equations (\ref{SUSDN cont eq spin lin par}) and the Euler equations (\ref{SUSDN Euler eq spin UP and D lin par}) do not contain the spin density $\delta S_{x}$, $\delta S_{y}$ and the magnetic field $\delta \textbf{B}$. So, they are decoupled from the spin evolution equations and the equations of field. The continuity equation and the Euler equation for the spin-up fermions are coupled to the continuity equation and the Euler equation for the spin-down fermions via the short range interaction. They lead to the two sound waves.

Substituting the velocity perturbation from equation (\ref{SUSDN cont eq spin lin par}) into equation (\ref{SUSDN Euler eq spin UP and D lin par}) we find a set of two equations for the perturbations of the partial concentrations.
A non-trivial solition of this set exists if the determinant of this set is equal to zero.
It gives the following equation
$$\biggl(\omega^{2}-(6\pi^{2}n_{0u})^{\frac{2}{3}}\frac{\hbar^{2}k^{2}}{3m^{2}}+\frac{\hbar^{2}k^{4}}{4m^{2}}\biggr)\times$$
\begin{equation}\label{SUSDN disp eq for sound waves}  \times\biggl(\omega^{2}-(6\pi^{2}n_{0d})^{\frac{2}{3}}\frac{\hbar^{2}k^{2}}{3m^{2}}+\frac{\hbar^{2}k^{4}}{4m^{2}}\biggr) =\biggl(\frac{gk^{2}}{m}\biggr)^{2}n_{0u}n_{0d}. \end{equation}

For the zero equilibrium spin polarization equation (\ref{SUSDN disp eq for sound waves}) simplifies to
\begin{equation}\label{SUSDN disp eq for sound waves   zero spin pol} \biggl(\omega^{2}-(3\pi^{2}n_{0})^{\frac{2}{3}}\frac{\hbar^{2}k^{2}}{3m^{2}}+\frac{\hbar^{2}k^{4}}{4m^{2}}\biggr) =\pm\frac{g n_{0}k^{2}}{2m}, \end{equation}
where we have included $n_{0u}=n_{0d}=n_{0}/2$.

The equations for the spin density are coupled to equations of field for $\delta B_{x}$, $\delta B_{y}$. They lead to the spin wave.

\begin{figure}
\includegraphics[width=8cm,angle=0]{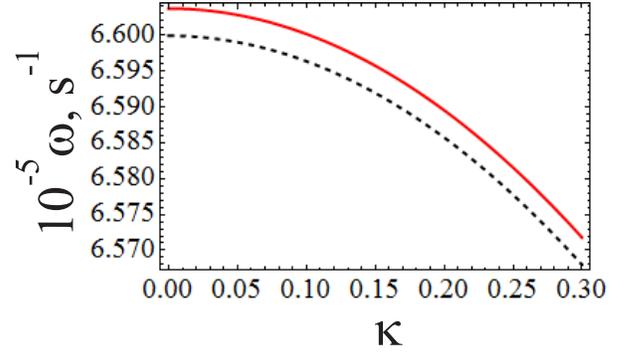}
\caption{\label{SEPAWOP F5} (Color online) The figure shows the dispersion dependencies of spin waves propagating parallel (\ref{SUSDN sw spectrum parallel}) (the upper curve) and perpendicular (\ref{SUSDN spectrum SW perp}) (the lower curve) to the external magnetic field at $\eta=0.3$, $\alpha=an_{0}^{1/3}=-0.4$, $\mu=2\mu_{B}$, $m=6$ u, $n_{0}=10^{14}$ cm$^{-3}$.}
\end{figure}

Spectrum of the spin waves arises as a result of evolution of $\delta S_{x}$ and $\delta S_{y}$ has the following form:
\begin{equation}\label{SUSDN sw spectrum parallel} \omega=\biggl|\Omega+\frac{g}{\hbar}\eta n_{0}-\textrm{w}-\frac{8\pi\mu^{2}}{\hbar}\eta n_{0}+\frac{\eta\varepsilon_{k}}{\hbar}\biggr|, \end{equation}
where
\begin{equation}\label{SUSDN} \textrm{w}=\frac{(6\pi^{2})^{\frac{2}{3}}\hbar}{m}(n_{0u}^{\frac{2}{3}}-n_{0d}^{\frac{2}{3}}) \end{equation}
is the characteristic frequency of the Fermi spin current, $\varepsilon_{k}=\frac{\hbar^{2} k^{2}}{2m}$ is the kinetic energy appearing from the quantum Bohm potential, $\Omega=2\mu B_{0}/\hbar$ is the cyclotron frequency, and $\eta=(n_{0u}-n_{0d})/n_{0}\in[-1,1]$ is the spin polarization.
The first and last terms in equation (\ref{SUSDN sw spectrum parallel}) are similar to the transverse spin wave mode obtained in Ref. \cite{Machida JPSJ 98} formula (14). Presence of a small magnetic field $B_{0}= 10^{-3}$ G makes the cyclotron frequency $\Omega$ comparable with other terms in (\ref{SUSDN sw spectrum parallel}). At the larger magnetic field the cyclotron term has the major contribution.

If $g>0$ and $\Omega+\eta gn_{0}/\hbar$ dominates over $w$ or $g<0$ and $\Omega$ dominates over $w+\eta \mid g\mid n_{0}/\hbar$, it gives us a positive constant $\Omega+\frac{g}{\hbar}\eta n_{0}-\textrm{w}-\frac{8\pi\mu^{2}}{\hbar}\eta n_{0}$ at $k=0$, we have increase of function $\omega(k)$ with the increase of the wave vector $k$ due to $\eta\varepsilon_{k}/\hbar$. If we have a negative constant at $k=0$ (it happens in the limits opposite to the described above) we have a decrease of the frequency $\omega$ at the increase of the wave vector $k$.

Solution (\ref{SUSDN sw spectrum parallel}) is presented in Fig. \ref{SEPAWOP F1} (a) for two different spin polarization. The frequency at the zero wave vector $\omega(k=0)$ is shifted from the cyclotron frequency by the short-range interaction (second term) and spin effects (third and fourth terms). The repulsive short-range interaction increases the frequency, while the spin effects decrease it.
In the considering range of parameters, the spin effects (the Fermi spin current dominates, while the fourth terms describing the spin-spin interaction is rather small) are dominate over the short-range interaction and the cyclotron frequency. Increasing the spin polarization we increase the frequency of spin wave.
The increase of the spin polarization increases all terms except the first term. However, the Fermi spin current grows faster since it is proportional to $[(1+\eta)^{2/3}-(1-\eta)^{2/3}]$ while the other terms are proportional to $\eta$.

Solution (\ref{SUSDN sw spectrum parallel}) is presented in Fig. \ref{SEPAWOP F1} (b) for three different masses of particles for a fixed spin polarization. Fixed spin polarization for a fixed scattering length at different masses appears at different magnetic fields. Hence, the cyclotron frequency is different for all curves. Change of mass itself changes the short range interaction term, the Fermi spin current term and the quantum Bohm potential. Altogether, effect of mass increase and corresponding change of magnetic field lead to the frequency decrease of spin waves.

Spectrum of the waves of concentrations $n_{u}$, $n_{d}$ (and the velocity fields $\textbf{v}_{u}$, $\textbf{v}_{d}$) appears from equations (\ref{SUSDN cont eq spin UP}) and (\ref{SUSDN Euler eq spin UP and D}):
\begin{equation}\label{SUSDN spectrum parallel longitud part} \omega^{2}=\frac{1}{2}\biggl[U_{u}^{2}+U_{d}^{2}\pm\sqrt{(U_{u}^{2}-U_{d}^{2})^{2}+4g^{2}\frac{n_{0u}n_{0d}}{m^{2}}}\biggr]k^{2}, \end{equation}
where $U_{s}^{2}=(6\pi^{2}n_{0s})^{\frac{2}{3}}\hbar^{2}/3m^{2}+\varepsilon_{k}/2m$.
Sign of the scattering length do not affect dispersion dependence (\ref{SUSDN spectrum parallel longitud part}).

If there is no spin polarization we have $n_{0u}=n_{0d}=n_{0}/2$ and $U_{0u}=U_{0d}$. It can be reached by a combination of a magnetic field and corresponding attractive short range interaction. Consequently equation (\ref{SUSDN spectrum parallel longitud part}) simplifies to $\omega^{2}=U^{2}k^{2} \pm gn_{0}k^{2}/2m$, with $U^{2}=(3\pi^{2}n_{0})^{\frac{2}{3}}\hbar^{2}/3m^{2}+\varepsilon_{k}/2m$. The upper sign in $\omega^{2}$ corresponds to $\delta n_{u}=-\delta n_{d}$ (the spin acoustic mode) and the lower sign corresponds to $\delta n_{u}=\delta n_{d}$ (the usual sound mode).

Using $\nabla\cdot \textbf{B}=0$ we find $\delta B_{z}k_{z}=0$, which gives $\delta B_{z}=0$. Thus, the magnetic field perturbation disappears from equations (\ref{SUSDN cont eq spin UP}) and (\ref{SUSDN Euler eq spin UP and D}). Therefore, the sound waves and the spin waves are described by the independent equations (in linear regime).

Obtained solution (\ref{SUSDN spectrum parallel longitud part}) is similar to the spin-electron acoustic wave found in Refs. \cite{Andreev PRE 15 SEAW}, and spin plasmon obtained for the two dimensional electron gas \cite{Ryan PRB 91}.

Equation (\ref{SUSDN spectrum parallel longitud part}) can be represented in the following form:
$$\omega=\frac{\hbar n_{0}^{\frac{2}{3}}}{m}\kappa \Biggl[\frac{(3\pi^{2})^{\frac{2}{3}}}{6}\biggl((1+\eta)^{\frac{2}{3}}+(1-\eta)^{\frac{2}{3}}\biggr)+\frac{\kappa^{2}}{4}$$
\begin{equation}\label{SUSDN spectrum parallel longitud part dimless} \pm\frac{\pi}{2}\sqrt{\biggl(\frac{\pi}{3}\biggr)^{\frac{2}{3}}\biggl((1+\eta)^{\frac{2}{3}}-(1-\eta)^{\frac{2}{3}}\biggr)^{2}+16(1-\eta^{2})\alpha^{2}}\Biggr]^{\frac{1}{2}} \end{equation}

The sound waves described by equation (\ref{SUSDN spectrum parallel longitud part}) are presented in Figs. \ref{SEPAWOP Facw1}-\ref{SEPAWOP Facw7}. Increasing the external magnetic field we increase the spin polarization. It increases the splitting of two branches of the sound waves Fig. \ref{SEPAWOP Facw1}. The upper wave is a modified sound wave existing even at the zero spin polarization. The lower branch is the spin acoustic wave.
Further increase of the spin polarization increases the splitting of two branches Fig. \ref{SEPAWOP Facw2}.
The spin acoustic wave spectrum is more affected by the spin polarization than the upper branch.
The increase of the spin polarization decreases the frequency of the spin acoustic wave Figs. \ref{SEPAWOP Facw1}, \ref{SEPAWOP Facw2}.
Large spin polarizations leads to the modification of form of the spin acoustic wave spectrum.
In this regime its almost linear spectrum starts at $k_{*}>0$, so $\omega(k_{*})=0$, as it is demonstrated in Fig. \ref{SEPAWOP Facw3}.

Next, consider two spin polarizations $\eta_{1}=0.1$ and $\eta=0.3$ at different scattering lengths $\alpha=an_{0}^{1/3}$.
At the increase of the scattering length up to $\alpha=0.5$ we see that the upper sound has almost same spectrum for different spin polarization.
However, the increase of the scattering length noticeable decreases spin acoustic wave spectrum. It can be seen from the comparison of the lower dashed lines in Fig. \ref{SEPAWOP Facw1} and Fig. \ref{SEPAWOP Facw4} found for $\eta=0.1$.
The nonlinearity of the spectrum becomes more noticeable for larger $\alpha$.
Same increase of the scattering length for $\eta=0.3$ gives more changes to the spectrum. In addition to the frequency decrease we find that spectrum starts at $k_{*}>0$. Hence, the increase of $\alpha$ and increase of $\eta$ changes the spin acoustic wave spectrum in the same way. Next, focus on the decrease of $\alpha$ in compare with its value used in Fig. \ref{SEPAWOP Facw1}.
It increases frequency of the spin acoustic wave and decreases frequency of the upper sound wave Fig. \ref{SEPAWOP Facw5}. Consequently, it decreases splitting between spectrums of sound waves Fig. \ref{SEPAWOP Facw5}.
These effects become more prominent at the further decrease of the scattering length (see Figs. \ref{SEPAWOP Facw6}, \ref{SEPAWOP Facw7}).

\subsection{Propagation perpendicular to the equilibrium spin polarization}

At the perpendicular propagation we choose the wave vector in the following form $\textbf{k}=\{k,0,0\}$.

Let us present simplified form of equations (\ref{SUSDN cont eq spin lin})-(\ref{SUSDN field eq lin}) existing in this regime:
\begin{equation}\label{SUSDN cont eq spin lin perp}
\omega\delta n_{s}=n_{0s}k\delta v_{sx}, \end{equation}
\begin{equation}\label{SUSDN Euler eq spin UP and D lin perp}-m\omega\delta v_{sx}+\frac{k \delta p_{s}}{n_{0s}} +\frac{\hbar^{2}k^{3}}{4m}\frac{\delta n_{s}}{n_{0s}}=-gk\delta n_{s'}\pm\mu k\delta B_{z},\end{equation}
$$-\imath\omega \delta S_{x}+\frac{(6\pi^{2})^{\frac{2}{3}}\hbar}{m}(n_{0u}^{\frac{2}{3}}-n_{0d}^{\frac{2}{3}})\delta S_{y}-\frac{\hbar k^{2}}{2m}\frac{S_{0}}{n_{0}}\delta S_{y}$$
\begin{equation}\label{SUSDN eq of magnetic moments evol lin perp x} -\frac{2\mu}{\hbar}\delta S_{y}B_{0}-\frac{g}{\hbar}(n_{0u}-n_{0d})\delta S_{y}=-\frac{8\pi\mu^{2}}{\hbar}S_{0}\delta S_{y},\end{equation}
$$-\imath\omega \delta S_{y}-\frac{(6\pi^{2})^{\frac{2}{3}}\hbar}{m}(n_{0u}^{\frac{2}{3}}-n_{0d}^{\frac{2}{3}})\delta S_{x}+\frac{\hbar k^{2}}{2m}\frac{S_{0}}{n_{0}}\delta S_{x}$$
\begin{equation}\label{SUSDN eq of magnetic moments evol lin perp y} +\frac{2\mu}{\hbar}\delta S_{x}B_{0}+\frac{g}{\hbar}(n_{0u}-n_{0d})\delta S_{x}=0,\end{equation}
and
\begin{equation}\label{SUSDN field eq lin perp}\begin{array}{c}
\textbf{e}_{y}(\delta B_{z}-4\pi\mu(\delta n_{u}-\delta n_{d}))-\textbf{e}_{z}(\delta B_{y}-4\pi\mu\delta S_{y})=0, \\
\delta B_{x}=0.
\end{array}
\end{equation}
Other projections of the velocity field are equal to zero: $\delta v_{sy}=\delta v_{sz}=0$. The right-hand side of the spin evolution equations (\ref{SUSDN eq of magnetic moments evol lin perp x}), (\ref{SUSDN eq of magnetic moments evol lin perp y}) contain perturbations of the magnetic field expressed via the spin density in accordance with the Maxwell equations (\ref{SUSDN field eq lin perp}).

The set of continuity and Euler equations (\ref{SUSDN cont eq spin lin perp}), (\ref{SUSDN Euler eq spin UP and D lin perp}) is coupled with equation of field for $\delta B_{z}$ via the last term in each Euler equation. As we can see $\delta B_{z}$ is proportional to $\delta n_{u}-\delta n_{d}$. Substituting the found form of $\delta B_{z}$ in the Euler equations (\ref{SUSDN Euler eq spin UP and D lin perp}) we find a closed set of equations for $\delta n_{u}$, $\delta n_{d}$, $\delta v_{ux}$, $\delta v_{dx}$ which is independent from the equations for the spin density $\delta S_{x}$, $\delta S_{y}$ (\ref{SUSDN eq of magnetic moments evol lin perp x}), (\ref{SUSDN eq of magnetic moments evol lin perp y}).
Equations for $\delta n_{u}$, $\delta v_{ux}$ are coupled to equations for $\delta n_{d}$, $\delta v_{dx}$ via two terms in the Euler equations (\ref{SUSDN Euler eq spin UP and D lin perp}): the short range interaction and the spin-spin interaction.
Evolution of concentrations and velocity fields leads to two sound modes.
Evolution of the spin density $\delta S_{x}$, $\delta S_{y}$ together with equations of field $\delta B_{x}=0$ and $\delta B_{y}=4\pi\mu\delta S_{y}$ leads to the spin wave.

Spectrum of the spin waves appears from equations (\ref{SUSDN eq of magnetic moments evol lin perp x}), (\ref{SUSDN eq of magnetic moments evol lin perp y}), (\ref{SUSDN field eq lin perp}) as follows:
$$\omega^{2}=\biggl(\Omega+\frac{g}{\hbar}\eta n_{0}-\textrm{w}+\frac{\eta\varepsilon_{k}}{\hbar}\biggr)\times$$
\begin{equation}\label{SUSDN spectrum SW perp} \times\biggl(\Omega+\frac{g}{\hbar}\eta n_{0}-\textrm{w}-\frac{8\pi\mu^{2}}{\hbar}\eta n_{0}+\frac{\eta\varepsilon_{k}}{\hbar}\biggr). \end{equation}
The spin wave dispersion dependence is different for parallel and perpendicular propagation. This difference appears due to the single term $\frac{8\pi\mu^{2}}{\hbar}\eta n_{0}$. This term $\frac{8\pi\mu^{2}}{\hbar}\eta n_{0}$ exists in one of two multipliers in formula (\ref{SUSDN spectrum SW perp}). While it exists in both multipliers at the parallel propagation (\ref{SUSDN sw spectrum parallel}). This difference is presented in Fig. \ref{SEPAWOP F5}.

Considering $\eta\sim0.1$, $B_{0}\sim10^{-3}$ G and larger, $n_{0}\sim10^{14}$ cm$^{-3}$, $g\sim10$ nm, we see that term $8\pi\eta\mu^{2}n_{0}/\hbar$ is the smallest term in equations (\ref{SUSDN sw spectrum parallel}) and (\ref{SUSDN spectrum SW perp}) (two or three order smaller then others). Hence, the spectrum of spin waves is almost isotropic. The difference of
spin wave spectrum for waves propagating in different directions is demonstrated in Fig. \ref{SEPAWOP F5}.

Spectrum of the sound waves is obtained as follows:
$$\omega^{2}=\frac{1}{2}k^{2}\Biggl[U_{u}^{2}+U_{d}^{2}-\Lambda^{2}$$
\begin{equation}\label{SUSDN sound perpend} \pm\sqrt{\biggl(U_{u}^{2}-U_{d}^{2}-\eta\Lambda^{2}\biggr)^{2}
+(g+4\pi\mu^{2})^{2}\frac{4n_{0u}n_{0d}}{m^{2}}}\Biggr], \end{equation}
where $\Lambda^{2}=4\pi\mu^{2} n_{0}/m$.
The effects of magnetic moment are relatively small in formula (\ref{SUSDN sound perpend}) for the considering parameters range. Therefore, the sound wave spectrum is almost isotropic.

\subsection{Perturbations at zero external magnetic field}

\begin{figure}
\includegraphics[width=8cm,angle=0]{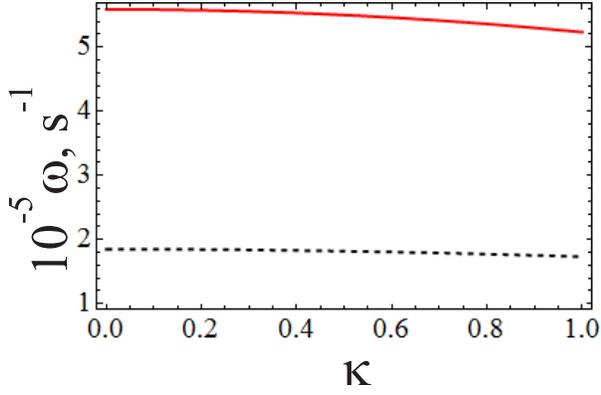}
\caption{\label{SEPAWOP F Spin_Waves_noMF_ATTR} (Color online)
The figure shows the dispersion dependence of spin waves propagating parallel to the equilibrium spin polarization caused by the short range interaction at the zero magnetic field. The frequency is plotted as a function of dimensionless wave vector $\kappa=k/n_{0}^{1/3}$. The figure shows the dispersion dependence for different spin polarization. The upper (lower) curve is obtained for $\eta=0.3$ ($\eta=0.1$), $n_{0}=10^{14}$ cm$^{-3}$, $\mu=2\mu_{B}$, $m=6$ u, $\alpha=-0.4$.}
\end{figure}

It is demonstrated in Sec. III A that the repulsive short range interaction between fermions with different spin projections can create an equilibrium spin polarization at the zero external magnetic field.
However, this regime is unstable to the Cooper pair formation. It is possible to consider the Fermi gas with an attractive short-range interaction and effective magnetic field $B_{eff}$ leading to the equilibrium spin polarization $\eta\neq0$.
Hence, we have equilibrium spin polarization creating the preferable direction $\textbf{S}_{0}=S_{0}\textbf{e}_{z}$. Consider perturbations of this equilibrium state. To do this we need to substitute $B_{0}=0$ in equations obtained above. We see that it does not change the analytical form of the sound wave solutions (\ref{SUSDN spectrum parallel longitud part}), (\ref{SUSDN sound perpend}), but it decreases the spin polarization existing in this equations. Substituting $B_{0}=0$ affects the spin wave solutions (\ref{SUSDN sw spectrum parallel}), (\ref{SUSDN spectrum SW perp}), where we shall put $\Omega=0$. Removing $\Omega$ changes the spin wave frequency as it is shown in Fig. \ref{SEPAWOP F Spin_Waves_noMF_ATTR}.
It increases frequency of spin waves, since we do not need to subtract the cyclotron frequency from the Fermi spin current term $w$.

Experimental realization of the described excitations is similar to the experimental realization of the collective excitations in spinor Bose-Einstein condensates \cite{Ho PRL 98}, \cite{Machida JPSJ 98}, \cite{Ueda AR CMP 11}.

\section{On a possibility of spin acoustic solitons}

If equilibrium spin polarization is caused by the radio-frequency
we have equilibrium effective magnetic field $B_{0}$.
Next, the spin evolution is proportional to the magnetic moment evolution.
Hence, it causes the appearance of the magnetic field.

The paper is focused on the nonlinear waves propagating parallel to
the direction of the equilibrium spin polarization
$\textbf{S}_{0}=S_{0}\textbf{e}_{z}$.
This direction of wave propagation leads to the following structure of the gradient operator: $\nabla=\{0,0,\partial_{z}\}$.

Consider perturbations of the magnetic field via the analysis of
the magneto-static Maxwell equations.
Equation $\nabla\cdot\textbf{B}=0$ is the first step of our analysis.
Substituting the gradient operator presented above in this equation find the following differential equation $\partial_{z}B_{z}=0$ which can be easily solved $B_{z}=c_{1}=B_{0}$.

The second step is the analysis of the equation $\nabla\times \textbf{B}=4\pi\mu\nabla\times \textbf{S}$.
It gives
$-\partial_{z}(B_{y}-4\pi\mu S_{y})\textbf{e}_{x}+\partial_{z}(B_{x}-4\pi\mu S_{x})\textbf{e}_{y}+0\textbf{e}_{z}=0$.
It leads to $B_{x}-4\pi\mu S_{x}=c_{2}$ and $B_{y}-4\pi\mu S_{y}=c_{3}$.
In equilibrium state we have $B_{x}=B_{y}=0$ and $S_{x}=S_{y}=0$.
Consequently, $c_{2}=c_{3}=0$.
Therefore, the magnetic field is found $B_{x}=4\pi\mu S_{x}$ and $B_{y}=4\pi\mu S_{y}$.
These relations between projections of the spin density and the magnetic field leads to the zero $z$-projection of the spin-torque entering the continuity equations: $T_{z}=[\textbf{B},\textbf{S}]_{z}=B_{x}S_{y}-B_{y}S_{x}=0$.

Present the modified set of hydrodynamic equations for the regime of perturbations propagating parallel to the equilibrium spin polarization:
\begin{equation}\label{SUSDN cont eq spin UP and d for non lin regime}
\partial_{t}n_{s}+\partial_{z}(n_{s}v_{sz})=0, \end{equation}
and
$$mn_{s}(\partial_{t}+v_{sz}\partial_{z})\textbf{v}_{s}+\delta^{\alpha z}\partial_{z} p_{s}$$
\begin{equation}\label{SUSDN Euler eq spin UP and D} -\frac{\hbar^{2}}{4m}n_{s}\delta^{z\alpha} \partial_{z}\Biggl(\frac{\partial_{z}^{2} n_{s}}{n_{s}}-\frac{(\partial_{z} n_{s})^{2}}{2n_{s}^{2}}\Biggr)=\textbf{F}_{g,s}+\textbf{F}_{S,s},\end{equation}
where $\textbf{F}_{g,s}$ is the force field of short-range interaction: $\textbf{F}_{g,u}=\{0,0,-gn_{u}\partial_{z} n_{d}\}$, $\textbf{F}_{g,d}=\{0,0,-gn_{d}\nabla n_{u}\}$, $\textbf{F}_{S,s}$ is the force field of spin-spin interaction, its explicit form arises as
$$F_{S,s}^{\alpha}=\pm\mu n_{s}\delta^{\alpha z}\partial_{z} B_{z} +\frac{\mu}{2}\delta^{\alpha z}(S_{x}\partial_{z} B_{x}+S_{y}\partial_{z} B_{y})$$
\begin{equation}\label{SUSDN force ss int} \pm 4\pi\frac{m\mu^{2}}{\hbar}(J_{(M)}^{x\alpha}S_{y}-J_{(M)}^{y\alpha}S_{x}),\end{equation}
with the following explicit form of the spin currents
\begin{equation}\label{SUSDN Spin current x} J_{(M)}^{\alpha\beta}=\frac{S^{\alpha}}{2}(v_{u}^{\beta}+v_{d}^{\beta})-\varepsilon^{\alpha\gamma z}\frac{\hbar\delta^{\beta z}}{4m} \biggl(\frac{\partial_{z} n_{u}}{n_{u}}-\frac{\partial_{z} n_{d}}{n_{d}}\biggr)S^{\gamma}. \end{equation}
Moreover, the first line in $F_{S,s}^{\alpha}$ is equal to zero:
\begin{equation}\label{SUSDN force ss int} F_{S,s}^{\alpha}=\pm 4\pi\frac{m\mu^{2}}{\hbar}(J_{(M)}^{x\alpha}S_{y}-J_{(M)}^{y\alpha}S_{x}).\end{equation}
Next, using explicit form of the spin-currents find
$$J_{(M)}^{x\alpha}S_{y}-J_{(M)}^{y\alpha}S_{x}$$
\begin{equation}\label{SUSDN} =-\frac{\hbar}{4m} \delta^{\alpha z} \biggl(\frac{\partial_{z} n_{u}}{n_{u}}-\frac{\partial_{z} n_{d}}{n_{d}}\biggr)(S_{x}^{2}+S_{y}^{2}).\end{equation}
It gives that force field $\textbf{F}_{S,s}$ has only z-projection $F_{S,s}^{\alpha}=\delta^{\alpha z}F_{S,sz}$:
\begin{equation}\label{SUSDN Fss simplified} F_{S,sz}=\mp \pi\mu^{2}\delta^{\alpha z} \biggl(\frac{\partial_{z} n_{u}}{n_{u}}-\frac{\partial_{z} n_{d}}{n_{d}}\biggr)(S_{x}^{2}+S_{y}^{2}).\end{equation}
Consequently, we obtain $n_{s}(\partial_{t}+v_{sz}\partial_{z})v_{sx}=0$ and $n_{s}(\partial_{t}+v_{sz}\partial_{z})v_{sy}=0$ while
$$mn_{s}(\partial_{t}+v_{sz}\partial_{z})v_{sz}+\partial_{z} p_{s}$$
\begin{equation}\label{SUSDN Euler eq spin UP and D one dim for non lin case} -\frac{\hbar^{2}}{4m}n_{s} \partial_{z}\Biggl(\frac{\partial_{z}^{2} n_{s}}{n_{s}}-\frac{(\partial_{z} n_{s})^{2}}{2n_{s}^{2}}\Biggr)=F_{g,sz}+F_{S,sz}.\end{equation}

It will be shown below that the spin-spin interaction force given by Eq. (\ref{SUSDN Fss simplified}) is the term of third order on the small parameter while our weakly nonlinear analysis includes terms up to the second order on the small parameter.
Hence, it can be neglected.
Therefore, there is no need to include the spin evolution equation in the following analysis.

\subsection{Small perturbations}

Equations (\ref{SUSDN cont eq spin UP and d for non lin regime}) and (\ref{SUSDN Euler eq spin UP and D one dim for non lin case}) are solved by
the perturbative method \cite{Washimi PRL 66}, \cite{Kalita PP 98}, \cite{Leblond JPB 08}.

We consider
$\xi=\varepsilon^{1/2}(z-Ut)$, and $\tau=\varepsilon^{3/2}Ut$,
where $U$ is the phase velocity of the wave.
The decomposition of the concentration and velocity field involves
a small parameter $\varepsilon$ in the following form:
$n_{s}=n_{0s}+\varepsilon n_{1s}+\varepsilon^{2}n_{2s}+...$, and
$v_{s}=\varepsilon v_{1s}+\varepsilon^{2}v_{2s}+...$.

Moreover, the spin density projections have similar decompositions $S_{x}=\varepsilon S_{1x}+...$ and $S_{y}=\varepsilon S_{1y}+...$.
These decompositions of the spin density make the spin-spin interaction force proportional to $\varepsilon^{7/2}$. So, it does not contribute in the two lowest orders $\varepsilon^{3/2}$ and $\varepsilon^{5/2}$ involved in our analysis.

From the lowest order on the parameter $\varepsilon$ of the hydrodynamic equations we find the phase velocity:
\begin{equation} \label{} U^{2}=\frac{1}{6}\biggl(v_{Fu}^{2}+v_{Fd}^{2}\pm\sqrt{(v_{Fd}^{2}-v_{Fu}^{2})^{2}+36\frac{g^{2}}{m^{2}}n_{0u}n_{0d}}\biggr). \end{equation}
Two phase velocities are found. It corresponds to the possibility of the existence of two solitons.

Considering terms proportional to $\varepsilon^{5/2}$, we find corresponding set of equations.
Representing all variables via $n_{1u}$, we find that $n_{1u}$ satisfies the Korteweg-de Vries equation:
\begin{equation} \label{SUSDN KdV eq sos}\partial_{\tau}n_{1u}
+p n_{1u}\partial_{\xi}n_{1u}+q \partial_{\xi}^{3}n_{1u}=0,\end{equation}
where
\begin{equation} \label{SUSDN q} q=-\frac{\hbar^{2}}{8 m^{2}U^{2}}, \end{equation}
and
\begin{widetext}
\begin{equation} \label{SUSDN p}
p=\frac{(3U^{2}-v_{Fd}^{2}/9)(U^{2}-v_{Fu}^{2}/3)^{2} +\frac{g}{m}n_{0d}(3U^{2}-v_{Fu}^{2}/9)(U^{2}-v_{Fd}^{2}/3)}{2U^{2} \frac{g}{m} n_{0u}n_{0d}(2U^{2}-(v_{Fu}^{2}+v_{Fd}^{2})/3)}. \end{equation}
\end{widetext}

Concentration $n_{1u}$ appearing as the solution of equation (\ref{TOIR Sol KdV eq sos}) can be presented in the following form
\begin{equation} \label{TOIR Sol wiev sol} n=n_{0}+\frac{2V\varepsilon}{p}\cdot sech^{2}\biggl(\sqrt{\frac{V}{q}}\frac{\eta}{2}\biggr), \end{equation}
where $\eta=\xi-V\tau$.

Parameter $q$ defines the width of the soliton. Hence, it should be positive for a soliton to exist.
However, parameter $q$ given by equation (\ref{SUSDN q}) is negative.
Hence, there is no soliton solution in this regime.

Let us mention that if there is no interaction $g=0$ then there is no soliton solution:
\begin{equation} \label{TOIR Sol KdV eq sos}\frac{2}{3}n_{0s}v_{Fs}^{2}\partial_{\tau}n_{1s}
+ \frac{8}{9}v_{Fs}^{2}n_{1s}\partial_{\xi}n_{1s}-\frac{\hbar^{2}}{4m^{2}} \partial_{\xi}^{3}n_{1s}=0.\end{equation}
It is correct for each species of fermions. As it is demonstrated above, the interspecies interaction does not change this picture.

We conclude that there is an analog of linear spin-electron acoustic wave (existing in the electron gas) in the gas of neutral fermions (this wave is called the spin acoustic wave).
However, there is no analog of the spin-electron acoustic soliton (existing in the electron gas) for neutral fermions with the short range interaction in the first order by the interaction radius.

\section{Conclusion}

A minimal coupling model of neutral weakly interacting spin-1/2 fermions with the short-range and spin-spin interactions has been presented in form of a non-linear Pauli equation with the spinor pressure term and two-fluid hydrodynamic equations. The presentation of the model in the hydrodynamic form explicitly shows the Fermi spin current (a part of the spin current caused by the Pauli blocking). Being in the partially polarized phase the spin-1/2 fermions show three collective excitations, two of them are sound waves, and the third wave is a spin wave.

The repulsive short range interaction leads to the spin polarization and the external magnetic field increases it.
Hence, the system shows an anisotropic behavior.
Described structure of the spectrum exist for both limit regimes of the wave propagation: parallel and perpendicular to the equilibrium spin polarization.
It has been shown that the Fermi spin current contributes in the spin wave spectrum only.
The Fermi spin current gives main contribution in the spin wave spectrum at the small magnetic field $B_{0}<0.01$.
The cyclotron frequency gives the main contribution at larger magnetic field.
Spectrum of the spin waves different at propagation parallel and perpendicular to the equilibrium spin polarization.
This difference by the spin-spin interaction.
However, the difference is rather small, so the spectrum is almost isotropic.
The increase of the spin polarization increases the frequency of the spin waves.

As it is mentioned above, there are two sound waves.
The upper sound wave exist due to the full Fermi pressure of fermions with both spin projections.
It can be found even in the single fluid model of spin-1/2 fermions.
The lower sound wave is the spin acoustic wave.
It exists due to relative motion of spin-up and spin-down fermions.

The upper sound wave increase its frequency at the increase of the spin polarization.
This effect is more pronounced at the small positive scattering length.
Change of spin acoustic wave frequency at the change of the spin polarization is larger then corresponding change of frequency of the upper sound wave.
However, the spin acoustic wave frequency decreases at the spin polarization increase.
The form of spin acoustic wave spectrum changes at the relatively large spin polarization and large scattering length.
In this regime the wave does not exist at small $k<k_{*}$, where $\omega(k_{*})=0$.
The spin acoustic wave frequency increases almost linearly at $k>k_{*}$.
The external magnetic field $B_{0}\geq0.01$ G creates full spin polarization of repulsive fermions.
In this regime, we have only one subspecies of fermions.
Hence, there is only one sound wave.

Overall, the equilibriums state and spectrum of small perturbations of weakly interacting spin-1/2 fermions are studied.



\begin{acknowledgements}
The work was supported by the Russian
Foundation for Basic Research (grant no. 16-32-00886) and the Dynasty foundation.
The author thanks Professor L. S. Kuz'menkov for useful discussions.
\end{acknowledgements}

\end{document}